# Bottom-gated epitaxial graphene suitable for half-integer quantum metrology ?

B. Jouault,[1] N. Camara,[1] B. Jabakhanji,[1] A. Caboni,[2] C. Consejo,[1] P. Godignon,[2] D. K. Maude,[3] J. Camassel[1]

[1]L2C– UMR 5650 Université Montpellier 2/CNRS, 34095 Montpellier cedex 5, France
[2]CNM-IMB-CSIC – Campus UAB 08193 Bellaterra, Barcelona, Spain
[3]Laboratoire National des Champs Magnétiques Intenses, CNRS-UJF-UPS-INSA, 38042 Grenoble, France

We demonstrate that the carrier concentration of epitaxial graphene devices grown on the C-face of a SiC substrate is efficiently modulated by a buried gate. The gate is fabricated via the implantation of nitrogen atoms in the SiC crystal, 200 nm below the surface, and works well at intermediate temperatures: 40K-80K. The Dirac point is observed at moderate gate voltages (1-20V) depending upon the surface preparation. For temperatures below 40K, the gate is inefficient as the buried channel is frozen out. However, the carrier concentration in graphene remains very close to the value set at T~ 40K. The absence of parallel conduction is evidenced by the observation of the half-integer quantum Hall effect at various concentrations at T~4K. These observations pave the way to a better understanding of intrinsic properties of epitaxial graphene and are promising for applications such as quantum metrology.

Because of the unique properties of the Dirac fermions, graphene-based devices are promising candidates to complement semiconductors in the future generations of microelectronic devices. Currently, the most promising techniques to produce graphene for industrial purposes seem to be either chemical vapor deposition on a metal[1,2] (CVD growth) or heating of a SiC wafer up to the so-called graphitization temperature[3] (epitaxial growth). In the first case, because the metallic film is electrically conducting, the graphene film needs to be transferred to a different substrate. In the second case, the graphene can be directly used on the insulating SiC surface. The main drawback of the epitaxial growth technique was the inability to control of the carrier density in situ. Recently, carrier density control has been achieved by various top-gate techniques[4][5]. While extremely promising for graphene-based electronics, these techniques are technologically cumbersome and hinder the use of the graphene surface for future applications such as gas or biomolecular sensors[6]. In addition, because of the coupling to the gate dielectric, top-gated devices obfuscate the role played by the SiC substrate in scattering processes and in the pinning of the Fermi level by interface states[7].

In recent work, epitaxial graphene on the Si-face of a SiC substrate was successfully bottom-gated[8]. Here, we follow this pioneering approach, with a number of simplifications and advances. Firstly, instead of the Si-face, we use as the starting material the C-face of a semi-insulating SiC substrate. In this case, no hydrogenation is required. Secondly, for the implantation, we use a single dose of nitrogen atoms, implanted at a rather low energy (160 keV). This implantation can be carried out on the full wafer scale, as done in this work, or more selectively, using a mask and growing graphene into the mask apertures. This should make the process easily transferable to microelectronics, requiring only industrial ion-implanters. Finally, keeping in mind that on the short term one of the most promising applications for epitaxial graphene is quantum metrology[9], we focus in this work on the low temperature physics and demonstrate that various quantum transport regimes can be achieved using this technology.

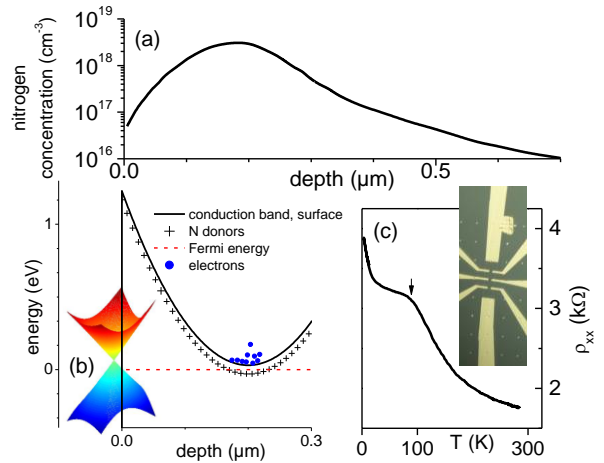

FIG. 1. (a) Simulation of the doping profile for nitrogen atoms introduced by implantation. Simulation was performed by the I2SiC in-house software from CNM Barcelona. The implantation energy is 160 keV. The maximum concentration of about $3 \times 10^{18}$ cm$^{-3}$ occurs ~ 200 nm below the surface. (b) Schematic of the graphene and SiC conduction bands at T ~ 50K. (c) Temperature dependence of the resistivity for one device. Below the inflexion point at T ~ 100 K (arrow), the graphene becomes more conductive than the implanted substrate. The sudden increase of the resistivity at $T < 20K$ is due to weak localization.

The implantation of nitrogen, the typical donor for SiC, was performed in a standard commercial accelerator on the C-face of a semi-insulating 6H-SiC on-axis substrate supplied by CREE Inc. An acceleration energy of 160 keV and a dose of 5 x 10$^{13}$cm$^{-2}$ in a EATON 4200 NV implanter was used. The calculated nitrogen profile, shown in Fig. 1(a), was obtained using in-house quasi-2D implantation software specifically developed for SiC. This tool is based on the Monte Carlo method within the binary collision approximation (BCA). The low nitrogen concentration at the SiC surface ensures the formation of an insulating barrier between the graphene and the conducting layer which acts as the gate. The conducting layer is located 100-200 nm below the surface, as indicated by the maximum of the N concentration (sketch in Fig. 1b).

The substrate was then annealed and graphitized in the same step; at 1700°C during 40 min, in a confined atmosphere, covering the substrate by a graphite cap, as described in previous work[10]. Graphene monolayer ribbons with typical widths of 5-10 µm and lengths up to 300 µm have been obtained by this method and contacted using e-beam lithography (see inset of Fig. 1c).

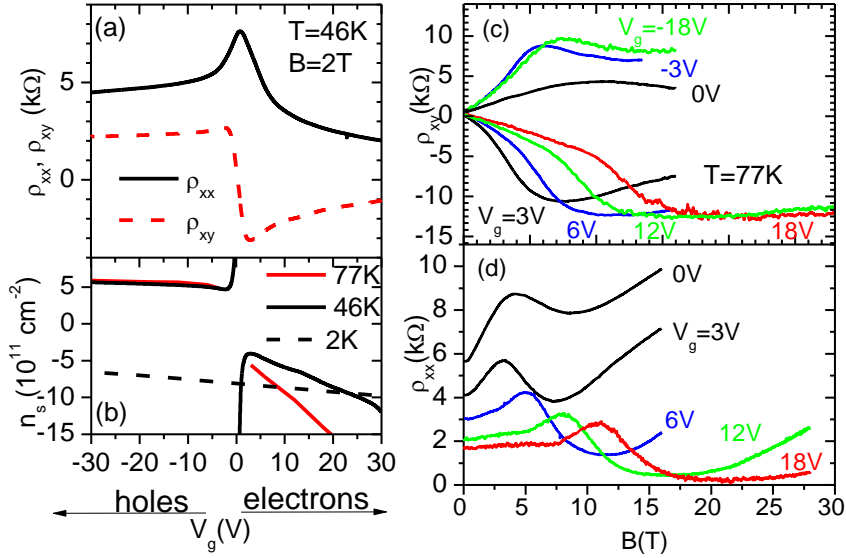

FIG. 2 (a) Longitudinal (solid line) and transverse (dashed line) resistivities $\rho_{xx}$ and $\rho_{xy}$ as a function of the gate voltage, at $T$=46K. The Dirac Point is observed at $V_g$ ~1V. (b) Estimated carrier concentration from the Hall measurements. The central divergence does not correspond to a homogeneous concentration, but rather reveals the coexistence of electrons and holes puddles. (c) Hall resistivity $\rho_{xy}$ as a function of the magnetic field up to 28 T, at a temperature of 77K and for different gate voltages from -18 V to 18 V. A transition from holes to electrons is observed around $V_g$=0 V. (d) longitudinal resistivity at various positive $V_g$, evidencing the shift of the Shubnikov-deHaas peaks (negative $V_g$ are not plotted for clarity).

Samples were cooled down to low temperatures using a cryostat equipped with a variable temperature insert. A magnetic field of up to 28 Tesla was applied perpendicular to the sample plane using a resistive magnet. The cooling procedure is crucial in order to be able to control the carrier density at low temperature. If the samples are cooled directly from 300 K to 1.5K (a typical temperature for metrological applications) with the gate floating or biased at 0V, then, at 1.5K, the carriers are holes with a concentration ~$10^{12}$ cm$^{-2}$, the implanted layer is frozen out and the carrier concentration cannot be modulated.

A better procedure consists, in a first step, to cool down to a temperature in the range 40-100K. In this range, transport properties are dominated by the graphene layer, while the buried channel remains conducting. At higher temperatures, current leaks through the insulating region become too important and mask the graphene conductivity. This is evidenced in Fig. 1c, where the inflexion point at T~100K in the device resistivity marks the beginning (with increasing temperature) of important current leaks through the buried channel. Fig. 2(a) shows the longitudinal and transverse resistivities $\rho_{xx}$ and $\rho_{xy}$ of the device in a magnetic field $B$=2T, at 46 K, in the working gate temperature range. We checked that the gate current was much lower that the source-drain current. Remarkably, the Dirac point is visible at $V_g$~ 1V, close to $V_g$=0V. The exact position varies from 1 to 20V and depends upon the surface preparation. The corresponding carrier concentration $n_s$= 1/ $e$ $\rho_{xy}$ is reported in Fig. 2(b), as well as for 2 other temperatures, which illustrate the loss of the gate efficiency at low $T$. By convention, $\rho_{xy}$>0 (<0) and $n_s$>0 (<0) correspond to holes (electrons). The divergence close to $V_g$=0V does not correspond to high carrier densities but is the usual signature of the coexistence of both electron and hole puddles.

To strengthen these points, Fig. 2c shows the Hall resistance of the device at a temperature of $T$=77K, for different gate voltages, as a function of magnetic field. The Hall resistance changes its sign around $V_g$=0V, confirming that carriers are holes for negative $V_g$, electrons for $V_g$>0. At high fields, $\rho_{xy}$ becomes quantized and saturates at $\rho_{xy}$ ~ -12 k$\Omega$ for positive gate voltages. The saturation is less pronounced for $V_g$<0V. Fig. 2(d) shows the longitudinal magnetoresistivity $\rho_{xx}(B)$ for various $V_g$. The shift of the $\rho_{xx}$ peak confirms the modulation of the carrier concentration. The quasi-cancellation of $\rho_{xx}$ at $V_g$~18V for $B$> 15T indicates that the quantum Hall regime is reached.

The capacitance of the gate can be calculated using $C$= $\Delta n_s$ $e$/ $\Delta V_g$. The asymmetry of the $n_s(V_g)$ curve in Fig. 2(b) with respect to the Dirac point indicates that the capacitance for the electrons and holes must be calculated separately. For electrons, we obtain $C$= 10nF/cm², 4nF/cm² and 0-0.8nF/cm² at 77K, 46K and 2K respectively. These values are significantly lower than the theoretical value $\varepsilon/d$~45nF/cm². The loss of the gate efficiency is partly due to the voltage drop in the implanted layer, whose resistance is very high at low temperature. Furthermore, the insulating region is very far from being a perfect Schottky barrier, as i) no increase of the capacitance at negative bias is observed; ii) the $I$-$V_g$ characteristic of the current leaks is mainly symmetric with $V_g$, and iii) a hysteretic behavior is detected in the device resistivity both in $V_g$ and $T$ below 40K, with very long relaxation times (~ hours at 20K). We conclude that the barrier characteristics are mainly governed by traps induced by implantation. The saturation of $n_s$ at 5x10$^{11}$ holes/cm² when $V_g$ < -10V (*i.e.* at an energy ~80 meV below the Dirac point) can be explained either by the presence of these traps, or by the presence of native interface states on the C-face of SiC. The presence of interface states on this face is still unclear[11 7].

Starting from a temperature T ~40-80K, the device can now be cooled down to 1.4 K at a fixed $V_g$. While the implanted layer becomes progressively frozen, the graphene monolayer retains its charge, as demonstrated in Fig. 4. Fig.4a, 4b and 4c show resistivities obtained at 1.4K, for which the cooling was initiated from 46K with different gate voltages (10V, 30V and -30V respectively). Fig. 3(a) shows a robust quantization of the plateau at 12.8kΩ from $B$ =7T to $B$> 28T. Carriers are electrons and quantization of both $\rho_{xx}$ and $\rho_{xy}$ persists for currents higher than 1μA (a proportionally large current density as the distance between the Hall probes is 1.7 μm). In Fig. 3(b), the electron concentration was slightly increased in order to reveal the quantized Hall plateau at 4 kΩ. In Fig. 3(c), carriers are holes at low fields, but the sudden decrease of the Hall resistance above 10T is a fingerprint of the coexistence of electrons and holes puddles at high fields[12]. No leakage current was detected at 1.5K and the resistance of the insulating region was greater than 100GΩ over the studied $V_g$ range. Precision of the Hall quantization (~0.1%) was limited by the measurement techniques (lock-in).

To summarize, we have demonstrated that the carrier concentration of epitaxial monolayer of graphene is efficiently modulated by an implanted buried gate. The Dirac point is observed at moderate gate voltages and at low temperatures, from 1.5K up to 77K, we observed well defined half integer quantum Hall effect. These observations, promising for low-T applications such as metrology, also pave the way to a better understanding of the role of the substrate on the transport properties of epitaxial graphene.

Part of this work has been supported by Euromagnet under contract GSC39-210, and by C'Nano GSO.

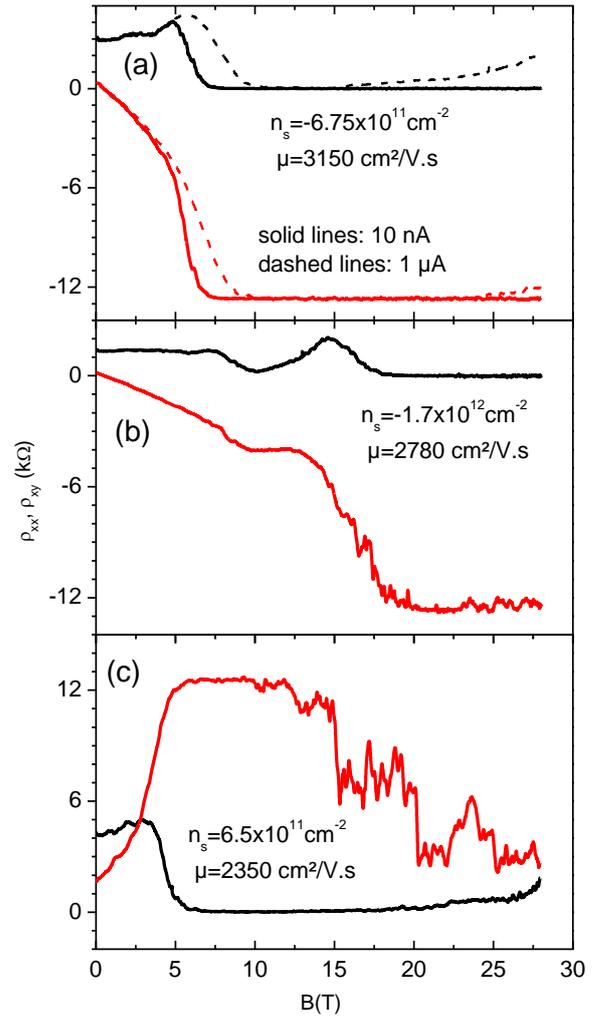

FIG. 3. Longitudinal (black) and transverse (red) resistivities $\rho_{xx}$ and $\rho_{xy}$ at different carrier concentrations. (a) electrons, showing a quantization between 7.5 and 28T; (a) electrons, higher concentration, where the plateau at 4 kΩ is visible in the transverse resistivity; (c) holes, with formation of electrons and holes puddles at high fields.